\documentclass[prb,twocolumn,floatfix,longbibliography, nofootinbib, noeprint]{revtex4-2}
\usepackage[utf8]{inputenc}
\usepackage[T1]{fontenc}
\usepackage{amsmath}
\usepackage{amssymb}
\usepackage{gensymb}
\usepackage{graphicx}
\usepackage{bm}
\usepackage{color}
\usepackage{hyperref}
\usepackage{upgreek}
\usepackage{scalerel}
\usepackage{pgffor}
\usepackage{pdfpages}
\usepackage{float}
\usepackage{appendix}
\usepackage{physics} 
\usepackage{lscape}   
\usepackage{silence}
\makeatletter
\AtBeginDocument{\let\LS@rot\@undefined}
\makeatother

\usepackage{soul}

\begin{document}

\title{Defect-induced band restructuring and length scales in twisted bilayer graphene}
\author{Lucas Baldo}
\affiliation{Department of Physics and Astronomy, Uppsala University, Box 516, S-751 20 Uppsala, Sweden}
\author{Tomas L\"othman}
\affiliation{Department of Physics and Astronomy, Uppsala University, Box 516, S-751 20 Uppsala, Sweden}
\author{Patric Holmvall}
\affiliation{Department of Physics and Astronomy, Uppsala University, Box 516, S-751 20 Uppsala, Sweden}
\author{Annica M. Black-Schaffer}
\affiliation{Department of Physics and Astronomy, Uppsala University, Box 516, S-751 20 Uppsala, Sweden}

\date{\today}

\begin{abstract}
	We investigate the effects of single, multiple, and extended defects in the form of non-magnetic impurities and vacancies in twisted bilayer graphene (TBG) at and away from the magic angle, using a fully atomistic model and focusing on the behavior of the flat low-energy moir\'e bands. For strong impurities and vacancies in the $AA$ region we find a complete removal of one of the four moir\'e bands, resulting in a significant depletion of the charge density in the $AA$ regions even at extremely low defect concentrations. We find similar results for other defect locations, with the exception of the least coordinated sites in the $AB$ region, where defects instead result in a peculiar band replacement process within the moir\'e bands. In the vacancy limit, this process yields a band structure misleadingly similar to the pristine case. Moreover, we show that triple point fermions (TPFs), which are the crossing of the Dirac point by a flat band, appearing for single, periodic, defects, are generally not preserved when adding extended or multiple defects, and thus likely not experimentally relevant. We further identify two universal length scales for defects, consisting of charge modulations on the atomic scale and on the moir\'e scale, illustrating the importance of both the atomic and moir\'e structures for  understanding TBG. We show that our conclusions hold beyond the magic angle and for fully isolated defects. In summary, our results demonstrate that the normal state of TBG and its moir\'e flat bands are extremely sensitive to both the location and strength of non-magnetic impurities and vacancies, which should have significant implications for any emergent ordered state.
\end{abstract}

\maketitle

\section{Introduction}
\label{sec:introduction}

Twisted bilayer graphene (TBG) has attracted considerable attention as both a versatile tunable experimental platform and a host of a plethora of ordered states \cite{Li2010, Cao2018, Kennes2018, Po2018b, Wu2018, Peltonen2018c, Cao2018d, Xie2019, Yankowitz2019, Lu2019, Jiang2019, Sharpe2019, Kerelsky2019, Choi2019, Serlin2020, Wu2020f, Christos2020, Balents2020, Chichinadze2020c, Andrei2020, Kang2021, Chou2021, Cao2021, Khalaf2021, Cea2021, Kim2022, Lothman2022, Wagner2022}, including both superconductivity and correlated insulating states \cite{Cao2018,Cao2018d,Cao2021}. Intriguingly, the wealth of ordered states in TBG is intimately connected to its remarkable unordered normal state electronic structure, where the Fermi velocity is suppressed with decreasing twist angle and even vanishes at so-called magic angles. The resulting flat dispersion has a large density of states and quenched kinetic energy that dramatically increases the importance of interactions and favors ordered states \cite{Peotta2015, Lothman2017}. As a consequence, around the magic angle, the ordered states of TBG depend crucially on the four spin-degenerate, emergent, low-energy, flat bands at the charge neutrality point (CNP), the moir\'e bands \cite{SuarezMorell2010, Bistritzer2010b, Lisi2021}. Among their distinctive properties, the moir\'e bands have been shown to have a topological obstruction from nonlocal symmetries that impose a lower bound on the localization of their associated Wannier orbitals \cite{Zou2018b, Po2018b}. Their unique energetics and spatial extent across the large emergent moir\'e pattern naturally prompts the question of how the moir\'e bands are impacted by explicitly local perturbations, such as atomic size lattice defects or impurities that are always present in any material. An answer to this question is important both in itself but also as it directly relates to the role of the moir\'e bands as a host of ordered states and as a versatile experimental probe.

When it comes to atomic size defects in single layer graphene, vacancies have been extensively studied and known to generate critically localized zero-energy states \cite{Pereira2006b}, exhibiting magnetism \cite{Yazyev2007b, Ugeda2010}. Although isolated vacancies are not thermodynamically stable due to high formation energy, several mechanisms can still lead to their formation \cite{Ban}. Focused electron beams, for example, allow for the creation of vacancies with close to atomic precision \cite{Rodriguez-Manzo2009}. Stone-Wales reconstructions and double vacancy structures on the other hand, have lower formation energies and are therefore even naturally ubiquitous \cite{Ban}. Moreover, adatoms and substitutional impurities constitute other common and well-studied types of defects in graphene \cite{Ban, Yang2018}. As such, defect studies have become an integral part of studying graphene. Atomic size lattice defects have also been studied in TBG. Many of these studies have only focused on the large-angle regime, including both more comprehensive studies \cite{Ulman2014b} and more focused studies into for example  fluorination process \cite{Muniz2013b, EkWeis2015b}, intercalation by lithium \cite{Larson2020b}, and more general charged defects \cite{Ramzan2022}. Other, somewhat related, works include the study of impurity-induced Friedel oscillations \cite{Lu2016b}, Raman spectroscopy of TBG samples with defects induced by ion beam irradiation \cite{Schmucker2015b}, and the study of vacancies and their migration \cite{Gong2017b}.
Moreover, vacancies in TBG have been found generate similarly localized states as in graphene, including also leading to Yu-Shiba-Rusinov (YSR) magnetically induced sub-gap states in the superconducting phase of magic-angle TBG \cite{Lopez-Bezanilla2019b}. Studies have also considered the interplay between defects and the pairing symmetries of the superconducting phase in magic-angle TBG \cite{Chen2019b, Yang2019b}. 
Finally, particularly relevant for this work is the finding of so-called triple point fermions (TPFs) \cite{Zhu2016, Bradlyn2016, Fulga2017, Xia2017, Wang2017, Cheung2018, Hutt2018, Kumar2019}, characterized by a triple band crossing, and associated valley polarization induced by single, weak, periodic impurities in TBG \cite{Ramires2019b}.

In this work we go beyond previous studies by providing a comprehensive investigation of the changes induced in the low-energy electronic structure of TBG by single, multiple, and extended atomic size lattice defects. By explicitly focusing on the impact low-energy electronic structure we both extract the inherent behavior of defects in TBG, which is likely notably different from graphene, and form a basis for understanding the implication of defects for emergent ordered states. To achieve this, we study TBG both near and away from the magic angle regime, for both weak and strong impurity strengths, and for both periodic and isolated defects. We primarily consider two types of lattice defects: non-magnetic potential impurities and vacancies, which model adsorbates, atomic replacements, and true vacancies and therefore effectively capture a wide range of different impurities and defects. We do so by employing fully atomistic tight-binding calculations including all carbon atoms and establish both the evolution of the band structure and the accompanied changes in the charge density as a function of impurity strength. 

In the case of periodic defects, we introduce one or more defects to the moir\'e unit cell, the emergent unit cell of TBG, thus preserving the translational invariance such that the band structure is still well defined. To access instead the effects of isolated defects, we effectively separate the defects by using supercells comprised of many moir\'e unit cells, still within a fully atomistic approach. The analysis of the periodic and isolated defects therefore complement each other with different experimental relevance. While understanding the isolated defect case is important in itself and for example, for quasiparticle interference studies which can probe symmetries of ordered states \cite{Hoffman2002, Hanke2012, Chi2014}, the periodic case has particular experimental and practical relevance based on contemporary impurity deposition techniques which enable the engineering of defect patterns \cite{DM1990, Custance2009}, as has been also recently illustrated by the synthesis of periodic molecular arrays on graphene with both atomic precision and tunable periodicity \cite{Lu2019b}. Additionally, the moir\'e unit cell itself defines an emergent periodic structure with a corresponding energy landscape that should intrinsically favor certain defect lattices, produce self-assembly of defects, or aid in the engineering of defect patterns. For instance, atomic hydrogen has been shown to preferentially adsorb following the moir\'e pattern produced between graphene and an Ir(111) substrate \cite{Balog2010}. The same moir\'e  pattern has similarly been shown to produce regular Ir and Pt clusters \cite{NDiaye2006,Franz2013,Linas2015}. For TBG, ab-initio methods have already shown that atomic hydrogen preferentially absorb with a higher binding energy to the $AA$ regions of the TBG moir\'e lattice \cite{Brihuega2018}. This suggests the possibility of structured patterning also of TBG and the engineering of periodic lattices of defects. 

Our results show that atomic size impurities and vacancies have a profound and special effect on the low-energy moir\'e band structure and thus directly on the properties of TBG. This is best illustrated by defects in the $AA$ region, where we find a complete removal of one entire moir\'e band from the low-energy band structure, even for only a single defect per moir\'e unit cell. At the magic angle this corresponds to an impurity concentration of only $\sim 0.01\%$, thus highlighting how extremely sensitive the low-energy electronic structure is to defects. This band removal results in a concomitant depletion of the $AA$ lattice regions, where the moir\'e bands are primarily concentrated \cite{Kang2018c, Koshino2018c}. With any interacting many-body ground state heavily dependent on the normal state low-energy moir\'e band structure, this defect-induced band removal will have severe consequences for the physics of TBG. We also observe an important dependence on the defect location. The defect-induced band removal also occurs for defects in the domain wall ($DW$) region and for the higher coordinated sites of the $AB$ region, while we instead find a band replacement occurring for defects in the least coordinated sites of the $AB$ region. Thus, even if the degeneracy of the low-energy moir\'e band structure is preserved in the last case, the properties of the bands are still completely altered. Moreover, we establish that the impurity strength necessary for an impurity to start behaving as a vacancy varies substantially for different impurity locations. Beyond the strong effect on the moir\'e bands we also find a localized defect state manifesting on the atomic length scale, which we trace back to the well-established defect state in monolayer graphene \cite{Pereira2006b, Ugeda2010}. The presence of this localized defect state is however  easily obscured by the depletion of the $AA$ region if the defect is also in this region. Thus there exist two length scales for defects in TBG: the atomic scale hosts a graphene-like defect state and the moir\'e scale controls the low-energy band structure and the change of charge density of the $AA$ regions. Furthermore, we find that the defect-induced triple degeneracy at the Dirac point, generating TPFs \cite{Zhu2016, Bradlyn2016, Fulga2017, Xia2017, Wang2017, Cheung2018, Hutt2018, Kumar2019}, earlier found for single defects in TBG \cite{Ramires2019b} is fundamentally not stable, but that the degeneracy is easily lifted with the introduction of either extended or multiple defects in the unit cell. We use degenerate perturbation theory to attribute this sensitivity of TPFs to an assumption of rank-$1$ perturbations, while more complex defect configurations generally violate such an assumption. 

Our results establish that the low-energy electronic structure of normal-state TBG changes drastically with the introduction of non-magnetic defects, changes that should be taken into account when considering the influence of the moir\'e bands on any electronic ordering achieved at low temperatures. As we report both band structure and charge density, our results are experimentally easily verified using for example angle-resolved photoemission spectroscopy (ARPES) or scanning tunneling spectroscopy (STS) and transport measurements, and they can also be directly extended to quasiparticle interference experiments. Our results can also be straightforwardly used to engineer altered band structures, including changing the number of moir\'e bands, even completely removing all moir\'e bands, or introducing flat bands at the Dirac point, and thereby possibly generating very different electronic orders at low temperatures. 

This work is organized in the following way. In Sec.~\ref{sec:model} we explain how we model TBG and its defects, including the observables used to examine the resulting electronic structure. Our results are presented in Sec.~\ref{sec:results}. In Sec.~\ref{sec:extreme-sensitivity} we show that the electronic structure of magic angle TBG is generally extremely sensitive to the presence of periodic defects. We then classify the defect locations into two distinct cases in Sec.~\ref{sec:defect-location}. In Sec.~\ref{sec:breaking-TPF} we complement earlier results by showing how the TPFs of TBG are broken by most realistic defects. In Sec.~\ref{sec:length-scales} we consider isolated defects, which allows us to identify two distinct length scales of defect behavior. We subsequently show how this behavior is not restricted to the magic angle and hence showcases universal physics of TBG. Finally, in Sec.~\ref{sec:conclusions} we draw our conclusions and propose ways in which our results could be tested experimentally. We also suggest implications of our work and pose questions for future investigation.

\section{Model and method}
\label{sec:model}

\begin{figure}[tb]
	\centering
	\includegraphics[width=\columnwidth]{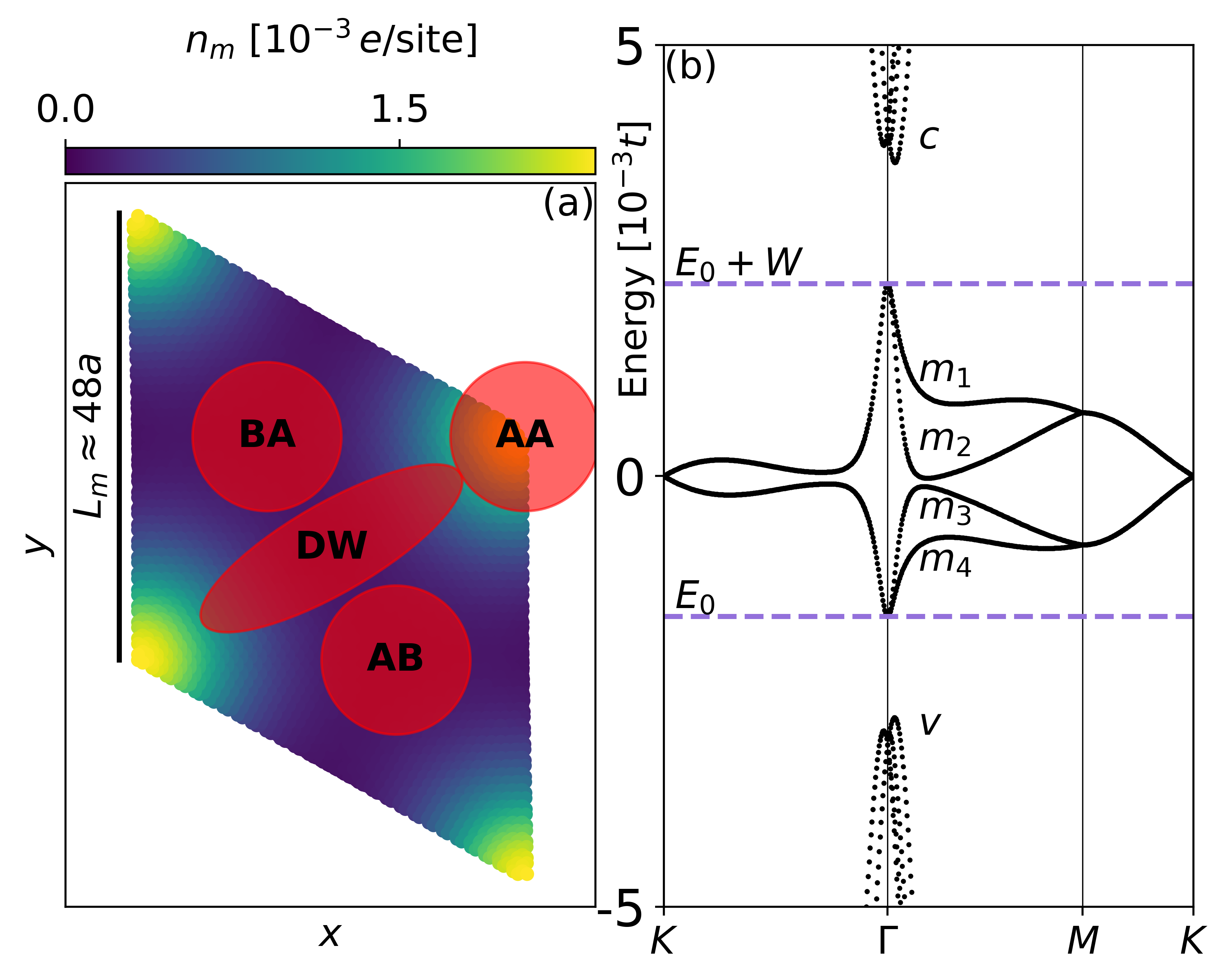}
	\caption{\label{fig:pristine} Charge density and band structure of pristine magic-angle TBG. (a) Charge density of the moir\'e bands of TBG, $n_{m} (\mathbf{x})$, with contributions from both layers shown. The $AA$, $AB$, and $DW$ regions of the moir\'e unit cell are schematically highlighted and the moir\'e length $L_m$ is marked. (b) Low-energy band structure of TBG, along high symmetry points: $K$, $\Gamma$, $M$. Horizontal dashed purple lines mark the moir\'e bandwidth, $W$. Moir\'e, conduction, and valence bands are labeled by $m_i$, $c$ and $v$, respectively. Here $\theta \approx 1.2\degree$ and we use a $12\times12$ $k$-point grid to compute (a).}
\end{figure}

TBG consists of two sheets of graphene stacked and rotated with respect to one another by a twist angle $\theta$. As a consequence, the lattice is modulated by an emerging length scale, creating a moir\'e pattern. For small twist angles and if the rotation axis goes through a graphene lattice site in both layers, then the relative alignment between the layers around this site remains, locally and approximately, as $AA$ layer stacking. Away from this site, the alignment transforms into $AB$ or $BA$ stacking depending on spatial direction. In between these directions, there is a transition region forming a domain wall ($DW$), where the stacking does not belong to either of these classifications. These regions are schematically depicted in Fig.~\ref{fig:pristine}(a) where we plot one moir\'e unit cell. Note that because of periodic boundary conditions the four corners in Fig.~\ref{fig:pristine}(a) are connected and thus the $AA$ region is split into the four corners of the moir\'e unit cell. The moir\'e unit cell forms a triangular lattice with lattice constant $L_m = a / (2 \sin(\theta/2))$ \cite{Shallcross2010}, also called the moir\'e length, where $a$ is the graphene lattice constant and $\theta$ is the twist angle. This lattice can be used to study the infinite system with atomic resolution, assuming appropriate commensuration conditions \cite{Shallcross2010}. We do this through a fully atomistic tight-binding model given by the Hamiltonian \cite{TramblydeLaissardiere2010, Moon2013}
\begin{equation}
	H_0(\mathbf{k}) = \sum_{i,j \in \mathbb{M}} t_{ij}(\mathbf{k}) c_{i \mathbf{k}}^\dagger c_{j \mathbf{k}},
	\label{eq:Hamiltonian}
\end{equation}
where $t_{ij}$ are hopping matrix elements under Bloch boundary conditions and the sum is taken over the sites of the moir\'e unit cell, $\mathbb{M}$. Near the magic angle, $\theta_m \approx 1.1 \degree$, this amounts to considering on the order of $10^4$ individual atoms. The operator $c^\dagger_{i \mathbf{k}}$ creates an electron on a site $i$ of the moir\'e unit cell. The layer, sublattice, and unit cell position of such a site are given by $l_i$, $s_i$, and $\bm{\rho}_i$, respectively. With $\bm{\delta}_{l_i}$ as the vector connecting $A$ and $B$ sites of layer $l_i$ and $d_0 = 3.35$~\r{A} the interlayer distance, the site positions are given by
\begin{equation}
	\mathbf{r}_i = \bm{\rho}_i + \delta_{s_i, B} \bm{\delta}_{l_i} + \delta_{l_i, 1} d_0 \hat{z},
	\label{eq:displacement-vector}
\end{equation}
where $\delta_{a,b}$ is the Kronecker delta. We thus opt to use a rigid lattice model, ignoring lattice relaxation effects. In terms of lattice effects, this relaxation has been shown to cause the $AA$ region to shrink, while in terms of the band structure it mainly rescales the magic angle and increases the gap between the moir\'e and the remote, conduction, and valence bands \cite{VanWijk2015a, Nam2017, Lucignano2019}. But, because the $AA$ area still remains a significant portion of the unit cell and, as we show, our main results are independent of twist angle, we do not expect lattice relaxation to strongly impact the effects of defects in TBG. Moreover, our implementation of the band structure, see below, achieves a finite gap isolating the moir\'e bands which is within experimentally measured bounds at the magic angle \cite{Lisi2021}, leading to a quantitatively correct capturing of the pristine moir\'e bands.

Using the above stated position vectors, the hopping elements of Eq.\eqref{eq:Hamiltonian} can be explicitly calculated. For intralayer hopping we only include next neighbor processes with $t_{ij}$ equal to the graphene hopping $t_\pi$. This is an approximation made in order to preserve the sparsity of the Hamiltonian matrix for computational efficiently. The result, compared to the full hopping model, is a rescaling of the twist angle, which we can simply compensate for to still achieve the magic angle, and an enhanced band gap isolating the moir\'e bands at the magic angle, which is beneficial when ignoring lattice relaxation as stated above. For the interlayer elements we use the Slater-Koster form \cite{Slater1954b, TramblydeLaissardiere2010}
\begin{equation} \label{eq:Slater-Koster}
	\begin{aligned}  
		t_{ij} (\mathbf{k}) = -\sum_{\mathbf{R}} e^{i \mathbf{R} \cdot \mathbf{k}} 
		& \left[ t_{\pi} e^{(a_c - r_{i j})/\lambda} (1- (\hat{r}_{i j} \cdot \hat{z})^2) \right. \\
		& \qquad \qquad \left. + t_{\sigma} e^{(d_0 - r_{i j})/\lambda} (\hat{r}_{i j} \cdot \hat{z})^2 \right],
	\end{aligned}
\end{equation}
where $\mathbf{r}_{i j} = \mathbf{r}_i - \mathbf{r}_j - \mathbf{R}$ is the displacement between the carbon sites $i$ and $j$ of unit cells connected by the lattice vector $\mathbf{R}$ of the moir\'e lattice. The sum over the lattice vectors $\mathbf{R}$ is important for sites near the edge of the unit cell, where the hopping occurs between different unit cells. The parameters $a_c = a/\sqrt{3}$ is the intralayer carbon to carbon distance. Other parameters are fixed according to the electronic structure of single and $AB$-stacked bilayer graphene \cite{TramblydeLaissardiere2010}, with $t_{\pi} = 2.7$~eV, out of plane hopping amplitude $t_{\sigma} = 0.48$~eV and overlap decay length $\lambda = 0.184 a_c$, which was shown to reproduce well the pristine band structure of TBG \cite{Lothman2022}. Because of the exponential form of the Slater-Koster terms, they become negligible for distant sites and we introduce a cutoff for $d > 6 a$ in order to preserve the sparsity of the Hamiltonian matrix, with no notable impact on the results. We also remark here that because there are no spin active terms in the problem we consider spinless fermions throughout the whole work. Hence, all observables can be thought of as of a single spin species and a multiplicative factor of $2$ leads to the values for the physical, spinful electrons.

From Eq.~\eqref{eq:Hamiltonian} we obtain the energy spectrum and eigenvectors of the system by diagonalization of the sparse Hamiltonian matrix using the eigenvalue solver PRIMME \cite{Stathopoulos2010}, focusing on the relevant energy regions near the charge neutrality point (CNP). The CNP, or half-filling, of the pristine system lies at the same energy as the Dirac point. The low-energy band structure is shown in Fig.~\ref{fig:pristine}(b) at a twist angle $\theta \approx 1.2\degree$, approximating the magic angle for our Hamiltonian. The four central low-energy bands are the so-called moir\'e bands, separated from the valence and conduction bands by a finite band gap and also with completely flat regions in the Brillouin zone at the magic angle \cite{LopesDosSantos2007, SuarezMorell2010, Shallcross2010, Bistritzer2010b}. Most of our calculations are done for $\theta \approx 1.2\degree$, modeling the magic angle regime, including the important finite band gap and flat band regions. More precisely, for the magic angle calculations, we use the parameters $(p,q)=(1,55)$ in the commensuration condition $\cos(\theta) = (3q^2 - p^2)/(3q^2 + p^2)$ \cite{Shallcross2010}, such that the moir\'e unit cell contains $9076$ atoms and the moir\'e length is $L_m \approx 47.6a$. However, as we later show, our findings hold for a much wider range of angles. We label the moir\'e bands from top to bottom as $m_1$, $m_2$, $m_3$, and $m_4$, and label the conduction and valence bands as $c$ and $v$, respectively. The energy range spanned by the moir\'e bands, from $E_0$ to $E_0 + W$, where $W$ is the moir\'e bandwidth, is from here on referred to as the \emph{moir\'e energy range}, $\mathbb{E}_m$. The boundaries of this range are outlined by horizontal dashed purple lines in Fig.~\ref{fig:pristine}(b), and in all later band structure plots, in order to create an explicit reference to the pristine case. By integrating the local density of states (LDOS) over the moir\'e energy range we obtain the charge density of the moir\'e bands,
\begin{align}
	n_{m} (\mathbf{x})
	& = e\int_{E_0}^{E_0+W} \text{LDOS}(\mathbf{x}, \epsilon) d\epsilon \nonumber \\
	& = \frac{e}{N_k} \sum_{E_{n, \mathbf{k}}\in\, \mathbb{E}_m} \vert \psi_{n, \mathbf{k}}(\mathbf{x}) \vert^2,
\end{align}
where $e$ is the electron charge, $N_k$ a normalization factor equal to the number of $k$-points sampled, and $\psi_{n, \mathbf{k}}$ are the eigenstates with energies $E_{n, \mathbf{k}}$. In each case we choose a grid density for reciprocal space sampling such that we observe a convergence of the main features in the charge densities. We show the moir\'e charge density for pristine TBG at $\theta \approx 1.2\degree$ in Fig.~\ref{fig:pristine}(a), where we see clearly  that the moir\'e bands are primarily localized in the $AA$ regions.

With the pristine tight-binding model established above, we now introduce defects into the lattice. Specifically, the defects we consider are non-magnetic potential impurities and vacancies. In order to introduce potential impurities we define a perturbing potential $v$ which enters the Hamiltonian as an onsite energy term on the affected sites 
\begin{equation} \label{eq:Hamiltonian-impurities}
	H_{I}(\mathbf{k}) =  \sum_{i \in \mathbb{M}} v_i c_{i \mathbf{k}}^\dagger c_{i \mathbf{k}}.
\end{equation}
For most of this work we focus on the simplest type of perturbing potential, which is that of a perfectly localized impurity with $v_i=E_I \delta_{i,d}$, where $E_I$ is the impurity strength and $d$ is the impurity site. In Sec.~\ref{sec:breaking-TPF} we additionally consider $v$ having a Gaussian profile around a central site, in order to investigate extended impurities. A vacancy can be introduced on the site $i$ by letting $v_i \rightarrow \infty$, such that this site effectively decouples from the rest of the lattice. For numerical stability, however, we use an equivalent approach of simply restricting the sum of Eq.~\eqref{eq:Hamiltonian} such that no hopping is allowed into or out of the vacancy sites $\mathbb{V}$,
\begin{equation} \label{eq:Hamiltonian-vacancies}
	H_{V}(\mathbf{k}) = \sum_{i,j \notin \mathbb{V}} t_{ij}(\mathbf{k}) c_{i \mathbf{k}}^\dagger c_{j \mathbf{k}}.
\end{equation}
For a perturbing potential $v$ or a set of vacancies $\mathbb{V}$ we obtain the energy spectrum of the total Hamiltonian $\Tilde{H}(\mathbf{k}) = H_{V}(\mathbf{k}) + H_{I}(\mathbf{k})$ near the CNP, where the moir\'e bands are located. Note that in the absence of vacancies, $H_V$ simply reduces to the pristine Hamiltonian $H_0$. These perturbing terms are effective models of both actual vacancies and chemisorbed adatoms in the case of $H_{V}$ and physisorbed adatoms in the case of $H_{I}$. By allowing the potential $v_i$ to have a finite spatial extent in $H_{I}$ and thereby creating extended impurities, we can even model larger physisorbed adatoms or even small molecules.

Because the creation and annihilation operators of Eq.~\eqref{eq:Hamiltonian} are of Bloch electrons, we effectively model a periodic lattice of defects, repeated in each moir\'e unit cell when solving $\Tilde{H}(\mathbf{k})$, even though the defect concentration is only 1 in $10^{4}$ for a single defect per unit moir\'e unit cell. We are also interested in the case of completely isolated defects. In order to study these within an atomistic model we turn to the use of supercells. In this approach we enlarge the unit cell of our lattice, by considering a supercell consisting of an $m \times n$ array of moir\'e unit cells, with only a single defect per such supercell. 
The periodicity of the defect is then that of the supercell and thus by choosing large enough supercells we can completely isolate the Bloch copies of the defects from one another, enabling us to study the isolated defect limit.

\section{Results}
\label{sec:results}

In order to perform an analysis of the effects that defects have on the low-energy electronic structure of TBG we begin with the most simple type of defect: a single-site impurity at site $d$, with the impurity potential given by $v_i = E_I \delta_{,id}$. In Sections \ref{sec:extreme-sensitivity} and \ref{sec:defect-location} we explore both the influence of the impurity strength, including the vacancy limit, and the defect location. We choose the defects to be always in the top layer since for single defects both layers are equivalent due to symmetry. Because of the sheer number of possible defect sites, we choose representative sites in the $AA$, $AB$, and $DW$ regions as candidates for the defect location. For the more computationally intensive calculations of the charge density, we focus on the vacancy limit and use the band structure results to guide our interpretations. Then in Section \ref{sec:breaking-TPF}  we turn our attention to putative TPFs at the Dirac point created by defects, and also extend our study to multiple defects per unit cell as well as defects with extended spread. Finally, in Section \ref{sec:length-scales} we study the length scale behavior of the effects of defects.
In Sections~\ref{sec:extreme-sensitivity}, \ref{sec:defect-location}, \ref{sec:breaking-TPF} we stay approximately at the magic angle using $\theta \approx 1.2$, while in Section~\ref{sec:length-scales} we study the behavior away from this regime.

\subsection{Extreme electronic structure sensitivity to atomic size lattice defects}
\label{sec:extreme-sensitivity}

\begin{figure}[tb]
	\centering
	\includegraphics[width=\columnwidth]{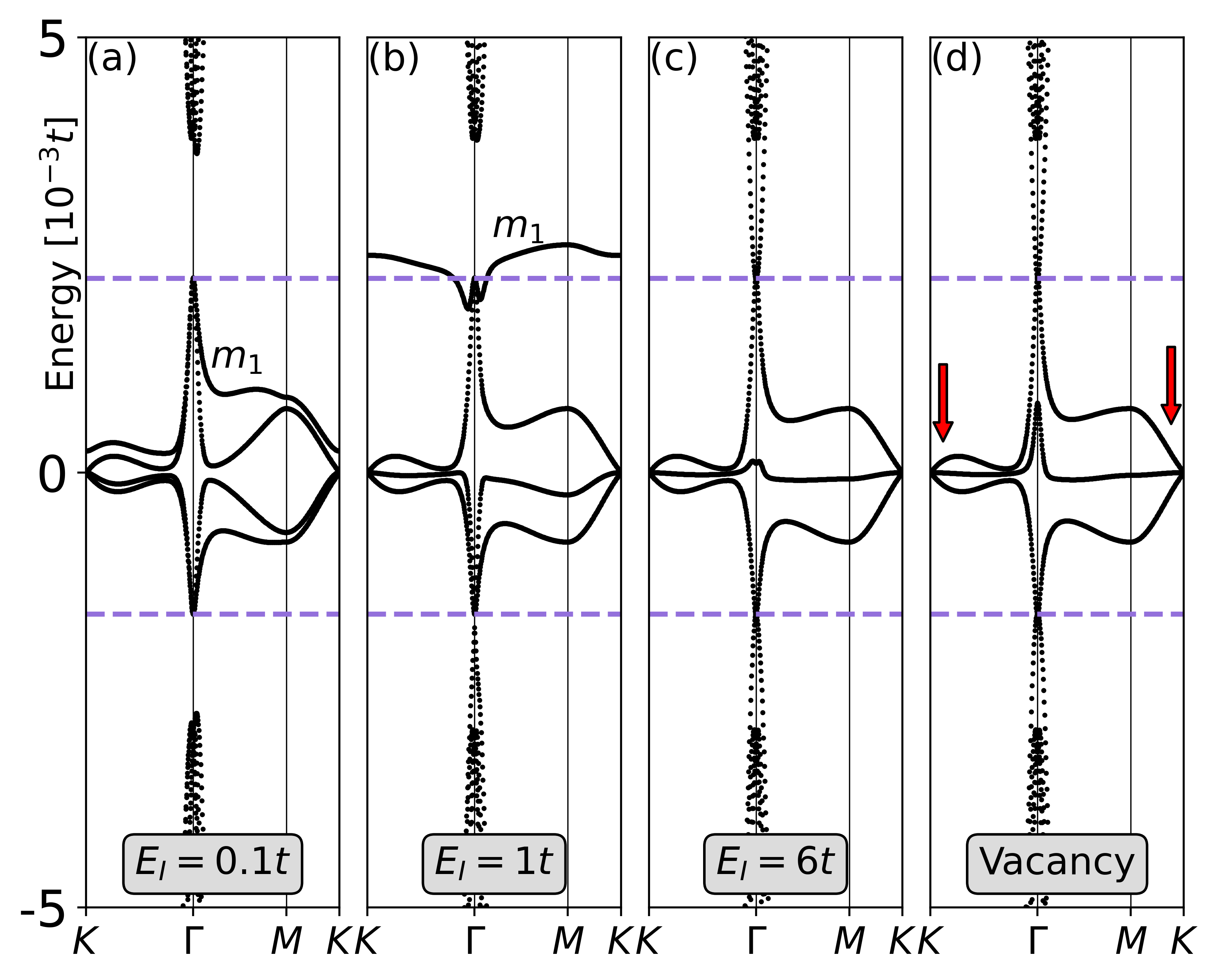}
	\caption{\label{fig:vacancy-limit} Effects on the low-energy spectrum of magic-angle TBG from a defect in the $AA$ region. (a-d) Band structure for potential impurities with strengths of $0.1t$, $1t$, $6t$, and a vacancy, respectively. Around $1t$ the moir\'e band $m_1$ leaves the moir\'e energy range, while the vacancy limit for the potential impurity is achieved around $6t$. Red arrows in (d) highlight the TPF. Here $\theta \approx 1.2\degree$.}
\end{figure}

Starting with a single defect per moir\'e unit cell, we find that for most defect locations the overall effect of an impurity or vacancy on the band structure is similar. The main features are most easily seen for the case of a defect in the $AA$ region, illustrated in Fig.~\ref{fig:vacancy-limit}. For an impurity strength up to $E_I =0.1t$, see Fig.~\ref{fig:vacancy-limit}(a), we find that the band structure changes very little even at the small energy scale of the moir\'e energy range. The most significant change seen is a breaking of the exact fourfold degeneracy of the Dirac point at $K$. However, as $E_I$ increases, the topmost of the four moir\'e bands, $m_1$, detaches from the others, except at the $\Gamma$ point, and lifts in energy. This band lifting is significant already at $E_I=1t$, see Fig.~\ref{fig:vacancy-limit}(b) where the $m_1$ band has already been removed almost entirely from the moir\'e energy range. This result shows how extremely sensitive the low-energy spectrum of TBG is with respect to defects, especially since the defect concentration here is only of the order of $\sim 0.01\%$. We note that this band removal behavior can be captured by band structure measurements, such as ARPES, or measurements of the density of states, such as STM or transport measurements, which would show a high peak from the relatively flat band $m_1$ at a much higher energy than the peaks due to the moir\'e bands in the pristine system. We also briefly note that $m_1$ never fully disconnects from the other moir\'e bands at $\Gamma$, while one of the valence bands $v$ lifts in energy just enough to touch $m_4$, closing the energy gap also from below and allowing $m_3$ to detach from $m_4$ at this point in the process. These degeneracies are discussed in more detail in Sec.~\ref{sec:defect-location}.

By further increasing the impurity strength we reach around $E_I=6t$ a behavior asymptotic in impurity strength, see Fig.~\ref{fig:vacancy-limit}(c) and Fig.~\ref{fig:vacancy-limit}(d) for a single vacancy. The resulting band structure contains three bands ($m_2,\,m_3,\,m_4$) within the moir\'e energy range, with the missing band ($m_1$) having acquired a parabolic character and joined the conduction bands. This results in a three-band moir\'e band structure that is not gapped neither from below or above. However, because $m_1$ and the highest lying valence band have both a very strong curvature, this band touching still results in a very small density of states when compared to the one from the leftover moir\'e bands.  Beyond the moir\'e bands only containing three bands we also find that the middle band (originally $m_3$) becomes much flatter compared to pristine TBG and lies essentially at the CNP. This directly exemplifies the possibility to engineer new flat band structures in magic-angle TBG by using impurities or vacancies.

\begin{figure}[tb]
	\centering
	\includegraphics[width=\columnwidth]{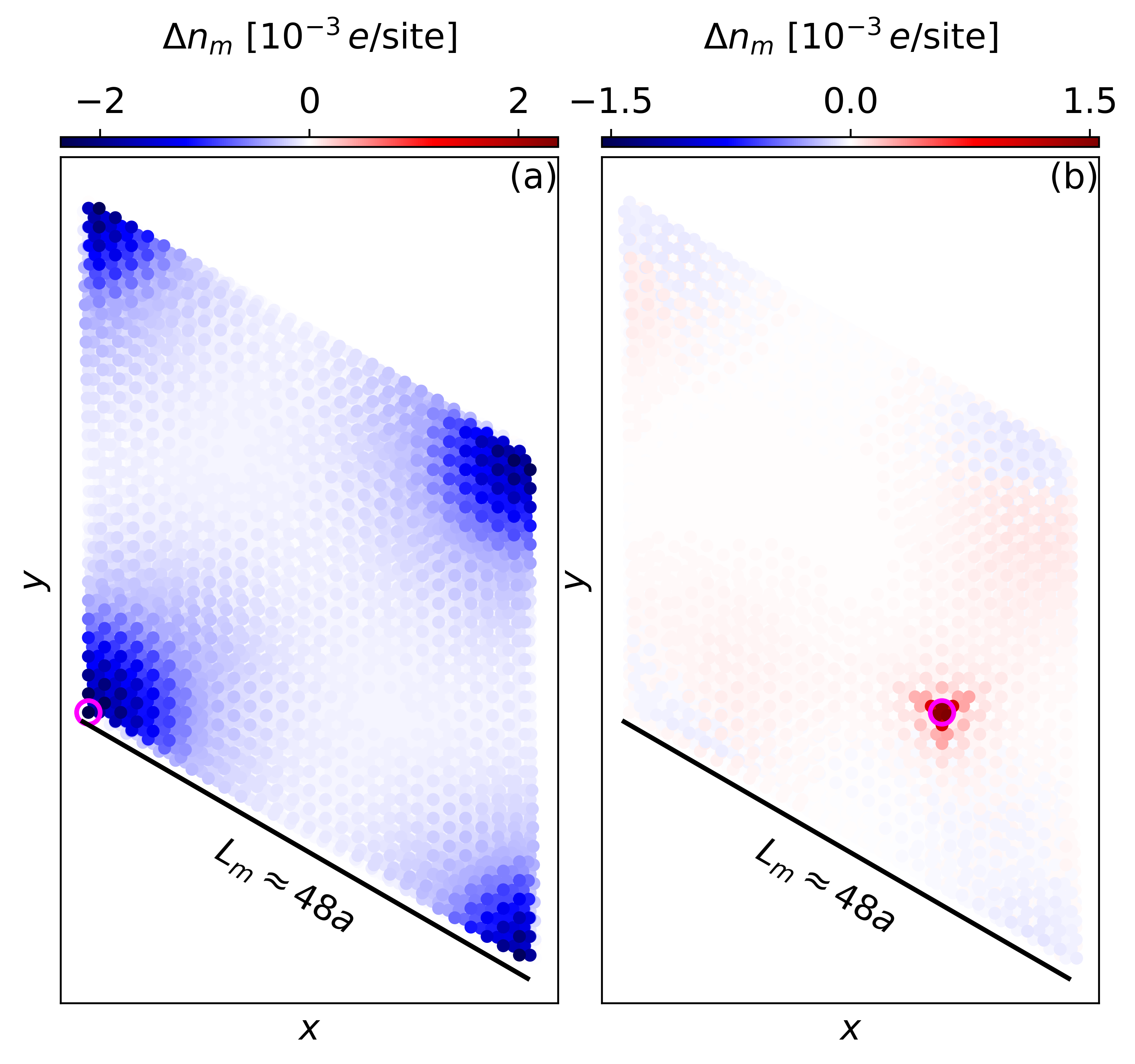}
	\caption{\label{fig:charge-density} Change in moir\'e charge density $\Delta n_m$ from pristine magic-angle TBG due to a vacancy. (a) Vacancy in the $AA$ region, showing a strong charge depletion of the $AA$ region, due to the removal of one of the moir\'e bands. (b) Vacancy in an $AB$-$LC$ site (defined in main text), where the main change in charge density is instead the graphene-like localized defect state. Pink circles mark the vacancy site and $k$-points were sampled in a $12\times12$ grid. In each panel the contributions from both layers are shown. Here $\theta \approx 1.2\degree$.}
\end{figure}

In Fig.~\ref{fig:charge-density}(a) we show the accompanying change in the moir\'e charge density, $\Delta n_m$, induced by the same vacancy as in Fig.~\ref{fig:vacancy-limit}(d). Each site in the unit cell is represented by colored dots, with the intensity of the color representing the magnitude of $\Delta n_m$. Blue (red) represents a decrease (increase) in charge density due to the vacancy, with the vacancy site encircled in magenta. Here the contributions from each layer are superimposed, with a finite transparency of the dots allowing for better visualization since the dots representing different sites often overlap, with the sites with greater magnitude in $\Delta n_m$ brought to the foreground.  We see that a vacancy in the $AA$ region induces a strong depletion of the charge density in the $AA$ region of the unit cell. This might at first seem counter-intuitive with respect to what is known about vacancies in graphene, where a vacancy is known to induce a zero-energy state centered on the vacancy, which causes a positive and localized change in charge density \cite{Pereira2006b, Ugeda2010}. Instead, the $AA$ depletion observed here has to be understood from the band structure of Fig.~\ref{fig:vacancy-limit}(d). Because one of the moir\'e bands, $m_1$, has been removed from the moir\'e energy range, and because these bands are localized in the $AA$ regions, see Fig.~\ref{fig:pristine}(a), the vacancy must cause a depletion of states over the entire $AA$ region, not just locally around the vacancy. A similar effect occurs for impurities with strengths above $1t$, since $m_1$ is already then removed from the moir\'e energy range, see Fig.~\ref{fig:vacancy-limit}(b). This extended effect of just single-site defects in TBG is in sharp contrast with how the same defect behaves in monolayer graphene, where the defect-induced state affects a much smaller region, and only locally around the defect. The connection between the depleted $AA$ regions and the lifting of $m_1$ can be further verified by integrating $\Delta n_m$ over the whole unit cell, where we find that there is $1e$ less charge in the integrated energy range, exactly corresponding to the removed moir\'e band. This overall charge depletion in the $AA$ region easily overshadows the monolayer graphene defect state in Fig.~\ref{fig:charge-density}(a), which we discuss in Sec.~\ref{sec:defect-location}.

\subsection{Role of defect location}
\label{sec:defect-location}

The results in Sec.~\ref{sec:extreme-sensitivity} were for a particular choice of a defect site in the $AA$ region. Now we also explore how the band structure and LDOS change for different defect locations. First of all, we find that different sites in each region within the same sublattice and in the same layer behave similarly to each other. Moreover, we find that the $AA$ and $DW$ regions show a sublattice symmetry with respect to the defect site, such that single defects in different sublattices lead to approximately the same band structure and charge density, assuming the sublattice components of the charge density are also switched $(n_m)_A \leftrightarrow (n_m)_B$. In all these cases we observe the removal of the $m_1$ band from the moir\'e energy range and the consequent depletion of the DOS in the $AA$ regions, as well as a migration of the $m_3$ band towards the CNP, where it becomes very flat in the vacancy limit. We note, however, that a larger impurity strength is needed in order to observe band structure changes for defects outside the $AA$ region, as expected since the moir\'e bands have much smaller presence outside the $AA$ region. In the $DW$ case we also observe that a valence band $v$ becomes flatter and rises in energy and, for some defect sites, even partially enters the moir\'e energy range. In this case, the $m_1$ band detaches at $\Gamma$, leading to an energy gap above the moir\'e bands. 

We next point out an observation we did not comment on in the last subsection. In the charge density for $AA$ defects, we observe that the charge depletion of the $AA$ region is reduced for sites immediately surrounding the defect that sit on the opposite sublattice to the defect site. This is a local effect, which we attribute to an induced, localized defect state akin to the one of monolayer graphene which appears on the sublattice opposite to the defect \cite{Pereira2006b, Ugeda2010}. In TBG, this state ends up partially canceling the change in charge density due to the removal of the $m_1$ band from the moir\'e energy range. Technically, for this defect location and at the magic angle, we cannot isolate the localized graphene-like defect behavior from the moir\'e depletion because the two effects are occurring in the same spatial location. However, this localized graphene-like defect state can be directly observed when the defect is in regions other than the $AA$ region. We find that it possesses a $C_3$ symmetry and decays almost within a few atomic sites, just as in monolayer graphene \cite{Pereira2006b, Ugeda2010}. For a $DW$ defect, in addition to the important $AA$ region depletion and the graphene-like localized defect state, we also observe a slight depletion of states within the $DW$ region itself, but it is contained to sites within the same sublattice as the defect site. For more details we refer to Appendix~\ref{app:defect-location}.

Moving on to single-site defects in the $AB$ region, we observe a behavior very similar to that of a $DW$ defect for half of the sites of the $AB$ region. More precisely, these are sites that have a neighboring site in the same position in the other layer. For this reason we call these sites higher-coordinated ($HC$) sites, while the other sites, which lie at the center of the carbon rings of the other layer, we refer to as the lower-coordinated ($LC$) sites. While a vacancy in an $AB$-$HC$ site leads to a very clear charge depletion of the $AA$ region, very similar to the behavior so far discussed (see Appendix~\ref{app:defect-location} for details), the introduction of a vacancy in an $AB$-$LC$ site leads to a drastically different charge density, as illustrated in Fig.~\ref{fig:charge-density}(b). In this case we find a much smaller $\Delta n_m$ in the $AA$ regions. Instead, the most pronounced feature is the graphene-like localized defect state with its $C_3$ rotational symmetry. A calculation of the band structure in this case yields a moir\'e structure with four bands, which explains why the charge density is similar to the pristine case. 

\begin{figure}[tb]
	\centering
	\includegraphics[width=\columnwidth]{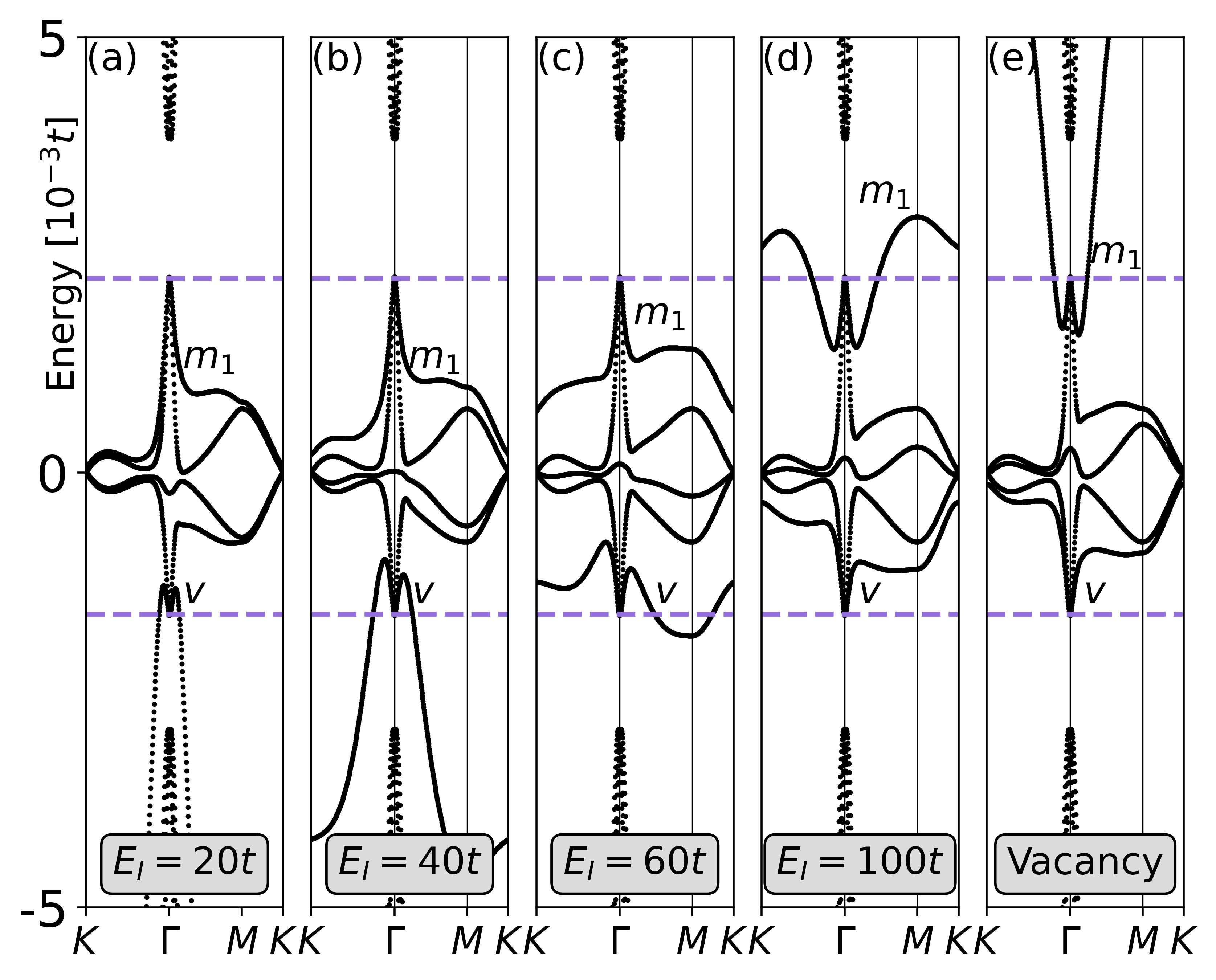}
	\caption{\label{fig:band-replacement} Band replacement process induced by a single-site impurity in an $AB$-$LC$ site. (a-e) Band structure for impurities with strengths $20t$, $40t$, $60t$, $100t$, and a vacancy, respectively. The vacancy limit is achieved for an impurity strength of around $500$t. Although initial and final band structures and their corresponding density of states are very similar, they do not contain the same bands, but a band replacement process takes place for increasing $E_I$, where the valence band $v$ enters the moir\'e energy range as the $m_1$ band is removed. Here $\theta \approx 1.2\degree$.}
\end{figure}

At first sight it might look as if TBG is thus insensitive to a vacancy on an $AB$-$LC$ site, given the largely unmodified charge density in $AA$ regions and the same number of bands, but this is not true. This precarious pitfall is revealed by the more careful analysis in Fig.~\ref{fig:band-replacement}, where we interpolate between the pristine case and the vacancy limit by tuning the impurity strength $E_I$. In Fig.~\ref{fig:band-replacement}(a) we show that even at the large value of $E_I=20t$ very little change from the pristine band structure is present among the moir\'e bands. The main difference is the attachment of a valence band $v$ to the moir\'e bands at $\Gamma$, closing the band gap. By further increasing $E_I$ up to $40t$, see Fig.~\ref{fig:band-replacement}(b), we find that this valence band $v$ starts a process of inverting its curvature. In this process, $v$ is initially outside of the moir\'e energy range, except at $\Gamma$, but as $E_I$ increases, the energy of $v$, especially around at the $M$ point, is brought into the moir\'e energy range, see Fig.~\ref{fig:band-replacement}(c). Then finally, with further increasing $E_I$, $v$ becomes less dispersive and completely joins the moir\'e bands, see Figs.~\ref{fig:band-replacement}(d-e). At the same time we observe the detaching and removal of the $m_1$ band from the moir\'e energy range. Thus, the similarity of the vacancy band structure to the pristine band structure is deceiving and the system has in fact undergone a \emph{band replacement} process in between: one band joins from the valence bands, while one of the pristine moir\'e band is lost to the conduction band in the vacancy limit. We notice here that the impurity strength necessary to drive this band replacement process is quite large, with the vacancy limit only being reached at around $E_I =500t$, to be compared to defects in the $AA$ region, where the vacancy limit is achieved already around $E_I =6t$. We further point out that the $AB$-$LC$ sites correspond to half of the sites in the $AB$ regions, which gives experimental relevance to these results.

Another interesting observation to be made about the role of the defect location is what happens to the degeneracies of the moir\'e bands at the $\Gamma$-point. In the pristine system the $m_1$ ($m_3$) and $m_2$ ($m_4$) bands are degenerate along the path from $K$ to $\Gamma$ (see Fig.~\ref{fig:pristine}). As defects are introduced, the degeneracy along this path is lifted as $m_1$ rises in energy. However, in the cases of $AA$ and $AB$-$LC$ defects shown in Figs.~\ref{fig:vacancy-limit} and \ref{fig:band-replacement}, the degeneracy remains at the $\Gamma$-point for all impurity strengths. This is because the wavefunctions of the moir\'e states at $\Gamma$ have nodes at these atomic sites. These nodes must be present because these states each constitute non-trivial irreducible representations (IRREPs) of the $C3$ symmetries of the lattice with rotation axes at the center of the $AA$, $AB$ or $BA$ regions. These IRREPs have eigenvalues $\exp(\pm i 2 \pi /3)$ for the generator of $C3$ and show a phase winding around the rotation axis. This leaves the phase at the central site undetermined and hence the wavefunction amplitude must exactly vanish there. As a consequence, the moir\'e states at $\Gamma$ are completely insensitive to perturbations added to these lattice sites, so the observed degeneracy is in these cases protected by the lattice symmetry. As the defect site is moved away from the center site of these regions, the $\Gamma$-point degeneracy is lifted, as illustrated by the cases of defects in $DW$ and $AB$-$HC$ sites (see Appendix~\ref{app:defect-location}). 
We also note that in the $AB$-$LC$ case of Fig.~\ref{fig:band-replacement} the $m_3$ and $m_4$ bands separate at $\Gamma$, but only after the latter is joined by a valence band, such that a twofold degeneracy is always present, although after this process the degeneracy is between $m_4$ and $v$.

Finally, we also point out that pristine TBG has an approximate valley symmetry, such that its bands can be classified according to a valley quantum number. Atomic size defects, however, lead in general to intervalley mixing, such that valley is no longer a good quantum number. In Ref.~\cite{Ramires2019b} the authors have shown that defects in different positions couple to the valleys in different ways. Defects in the AB and BA regions preserve valley polarization, such that the TPF structure shows an intact Dirac cone from one valley and a split Dirac cone from the other valley. On the other hand, defects in the AA and DW regions lead to valley-unpolarized states, which can be understood as a consequence of intervalley mixing due to the defect. Due to these extensive intervalley processes, we refrain from further discussions regarding the valley quantum number.

\subsection{Destroying triple point fermions}
\label{sec:breaking-TPF}

For all the defects we considered so far, a triple degeneracy of the moir\'e bands exists at the Dirac points $\mathbf{K}$, $\mathbf{K}'$, see e.g.~red arrows in Fig.~\ref{fig:vacancy-limit}(d). This gives rise to a TPF \cite{Zhu2016, Bradlyn2016, Fulga2017, Xia2017, Wang2017, Cheung2018, Hutt2018, Kumar2019, Ramires2019b}, since there is a Dirac spectrum crossed by a flat band, giving rise to a triply degenerate Dirac point. This is in contrast with the pristine TBG case, where, because of the valley degree of freedom, there are two degenerate Dirac cones at $\mathbf{K}$, $\mathbf{K}'$, leading to a fourfold degeneracy of the Dirac point. Hence, the TPF scenario represents a reduction of the degeneracy of the Dirac point. The presence of TPFs in TBG has already been discussed in Ref.~\cite{Ramires2019b}, where their robustness with respect to the impurity strength has been highlighted. Here we extend these results by showing that this robustness actually relies on the assumption of a single-site perturbing potential. In fact, we show that the TPF is split upon the introduction of either multiple defects or an extended impurity.

Let us begin by investigating the three-fold degeneracy in more detail, focusing on the spectrum near $\mathbf{K}$, as a similar argument holds for the other inequivalent Brillouin zone corner $\mathbf{K}'$. In the pristine case there is an exact fourfold degeneracy at this point due to Dirac cones from the two so-called valleys of TBG, which originate from the two layers. Moreover, because the conduction and valence bands are strongly dispersive in comparison with the moir\'e bands, there is a sizable gap $\Delta E_\mathbf{K}$ in this region of reciprocal space, much larger than the energy splitting of the moir\'e bands $E_{m_1}(\mathbf{k}) - E_{m_4}(\mathbf{k}) = 2 v_F |\mathbf{k} - \mathbf{K}|$ near $\mathbf{K}$, where $\mathbf{k}$ is a lattice wave vector near $\mathbf{K}$ and $v_F$ is the Fermi velocity corresponding to the slope of the Dirac cones of TBG. Because of this, it is possible to treat the effect of a weak perturbation on the moir\'e bands around $\mathbf{K}$ by projecting it into the moir\'e band subspace. Particularly at the Dirac point, $\mathbf{k} = \mathbf{K}$, because of its fourfold degeneracy, we need to use degenerate perturbation theory. 

With the above considerations, the energies of the moir\'e bands can be approximated by projecting the perturbed single particle Hamiltonian $h = h_0 + h_I$ into the moir\'e subspace, where $h_0$ and $h_I$ are the single particle versions of Eqs.~\eqref{eq:Hamiltonian} and \eqref{eq:Hamiltonian-impurities} evaluated at $\mathbf{K}$. Letting $P_m \equiv \sum_i | m_i \rangle \langle m_i |$ be the projector into this subspace, where $|m_i \rangle$ are the degenerate eigenstates of the moir\'e bands at $\mathbf{K}$, the projected Hamiltonian becomes
\begin{align}
	\Tilde{h}
	&= P_m h P_m \nonumber \\
	&= P_m h_0  P_m + P_m h_I  P_m \nonumber \\
	&= E_\mathbf{K} P_m + \Tilde{h}_I,
\end{align}
where we use the fact that the pristine moir\'e bands are fourfold degenerate at the Dirac point, with energy $E_\mathbf{K}$ and further define $\Tilde{h}_I \equiv P_m h_I P_m$. We next recall that a rank-$1$ operator $O$ has only one non-zero eigenvalue $\epsilon$ and can be written as an outer product $O = \epsilon | \epsilon \rangle \langle \epsilon |$. Using this, we note that for a single-site impurity at site $\mathbf{x}_0 $, $h_I = E_I |\mathbf{x}_0\rangle \langle \mathbf{x}_0 |$, both $h_I$ and $\Tilde{h}_I$ are rank-$1$ operators. The proof for $h_I$ follows directly from its definition. For $\Tilde{h}_I$ we have
\begin{align}
	\Tilde{h}_I 
	&= E_I \sum_{ij} | m_i \rangle \langle m_i |\mathbf{x}_0\rangle \langle \mathbf{x}_0 | m_j \rangle \langle m_j | \nonumber \\
	&= E_I \sum_{ij} \psi_{m_i}^*(\mathbf{x}_0) \psi_{m_j}(\mathbf{x}_0)  | m_i \rangle \langle m_j | \nonumber \\
	&= E_I  \left( \sum_{i} \psi_{m_i}^*(\mathbf{x}_0) | m_i \rangle \right) \left( \sum_{j} \psi_{m_j}(\mathbf{x}_0) \langle m_j | \right) \nonumber \\
	&= E_I  | E_I \rangle \langle E_I |,
\end{align}
where $\psi_{m_i}(\mathbf{x}) = \langle \mathbf{x} | \psi_{m_i} \rangle$ is the wavefunction of the moir\'e band $m_i$ at $\mathbf{K}$ and in the last line we simply defined $| E_I \rangle \equiv  \sum_{i} \psi_{m_i}^*(\mathbf{x}_0) | m_i \rangle$. Thus the operator $\Tilde{h}_I$ has a single non-zero eigenvalue $E_I$ and is a rank-1 operator. As a consequence $\Tilde{h}$ has three degenerate eigenvalues $E_\mathbf{K}$ and another eigenvalue $E_\mathbf{K} + E_I$. This means that a single-site impurity reduces the degeneracy at the Dirac point from fourfold to threefold, leading to the formation of a TPF, exactly as earlier predicted \cite{Ramires2019b}.

\begin{figure}[tb]
	\centering
	\includegraphics[width=\columnwidth]{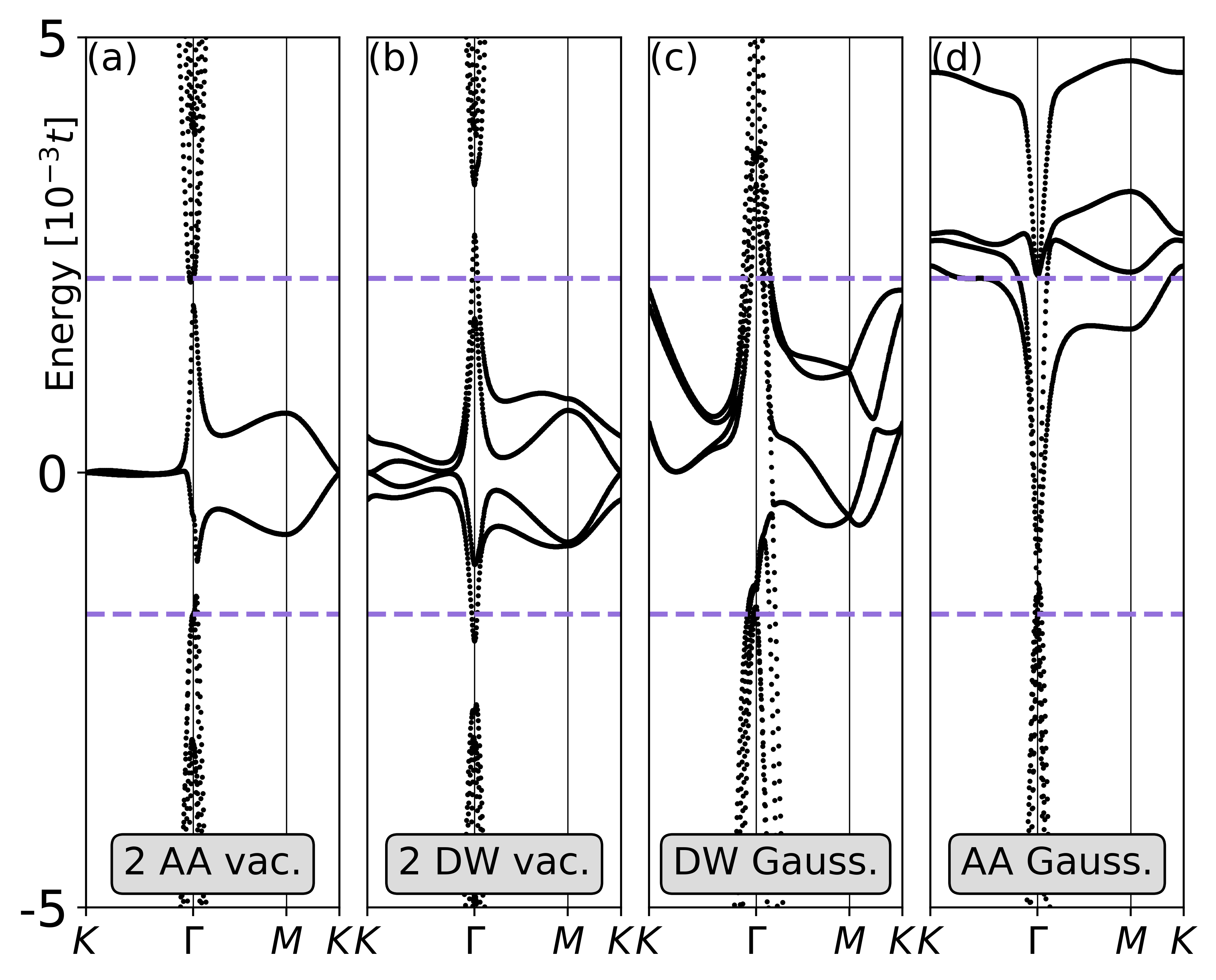}
	\caption{\label{fig:TPF-breaking} Removal of the TPF of TBG. (a-b) Two vacancies per unit cell in the $AA$ (a) and  $DW$ (b) regions. (c-d) Extended Gaussian impurity centered in the $DW$ (c) and $AA$ (d) regions. Gaussian spread of $\sigma=0.8a$ (c) and $\sigma=0.5a$ (d) with a cutoff of $2.5a$. In all cases (a-d) the TPF is split. Here $\theta \approx 1.2\degree$.}
\end{figure}

Next, let us consider two distinct, but still weak and perfectly localized impurities, $h_I^{(1)}$ and $h_I^{(2)}$. What we are interested in is whether the TPF found above remains for the total perturbation $h_{I}^{(t)} = h_I^{(1)} + h_I^{(2)}$. In order to do this, we again investigate the rank of the projected total perturbation, $\Tilde{h}_{I}^{(t)} = P_m h_{I}^{(t)} P_m$. We can quickly verify that the unprojected perturbation $h_{I}^{(t)}$ must be of a higher rank. One way to do this is to tentatively assume that $\Tilde{h}_{I}^{(t)}$ is rank-$1$. This means that it can be expanded as an outer product. By definition, this requires that there exist $\alpha_i$ such that
\begin{align}
	\Tilde{h}_{I}^{(t)}
	&= \sum_{i,j} | m_i \rangle \langle m_i | (h_I^{(1)} + h_I^{(2)}) | m_j \rangle \langle m_j | \nonumber \\
	&= \sum_{i,j} \left[ (h_I^{(1)})_{ij} + (h_I^{(2)})_{ij}  \right] | m_i \rangle \langle m_j | \nonumber \\
	&= \sum_{i,j} \alpha_i^* \alpha_j | m_i \rangle \langle m_j |,
\end{align}
where $h_{ij} = \langle m_i | h | m_j \rangle$ are the matrix elements of an operator $h$ in the basis of the moir\'e bands $| m_j \rangle$ and in the last line we used the definition of an outer product. For this last equality to hold $\Tilde{h}_{I}^{(t)}$ must be separable, that is, there must exist $\alpha_i$ such that
\begin{align}
	E_I^{(1)}\psi_{m_i}^*(\mathbf{x}_1) \psi_{m_j}(\mathbf{x}_1) + E_I^{(2)} \psi_{m_i}^*(\mathbf{x}_2) \psi_{m_j}(\mathbf{x}_2) 
	&= \alpha_i^* \alpha_j.
\end{align}
This constraint is however too restrictive and is, in the general case, not satisfied by any $\alpha_i$. Thus, in general, $h_{I}^{(t)}$ cannot be rank-$1$ and thus the introduction of a second weak impurity in the unit cell leads to a further change in the degree of degeneracy at the Dirac point and a subsequent splitting of the TPF. Moreover, although the argument above does not hold for arbitrarily large impurity strengths $E_I$, since we assumed $E_I \ll \Delta E_{\mathbf{K}}$ at the start, we verify through extensive numerical calculations that the conclusion that $h_{I}^{(t)}$ is not rank-$1$ and thus that multiple defects further reduce the degeneracy of the Dirac point still holds even in the vacancy limit. As an example, in Fig.~\ref{fig:TPF-breaking}(a) we show the band structure for the simple case of two nearby vacancies in the $AA$ region at sites of opposite sublattice in the same layer. In this case we observe the removal of another moir\'e band from the moir\'e energy range as compared to the single vacancy case. This clearly splits the TPF by lifting the needed degeneracy, as there are now only two bands degenerate at $\mathbf{K}$. We verify that this splitting occurs for many other defect locations, including for vacancies in different parts of the unit cell, such as one in the $AA$ region and the other in the $DW$ region. The only exception we find so far, where the TPF survives the inclusion of more than one single-site impurity, is tied to the lower-coordinated sites in the $AB$ region, $AB$-$LC$, where the band replacement process discussed in see Sec.~\ref{sec:defect-location} can in fact restore the TPF.

Fig.~\ref{fig:TPF-breaking}(a) also illustrates another interesting feature, namely that the effect of introducing two vacancies in the $AA$ region is an additive process of the effects of the single vacancies, in the sense that each vacancy is responsible for removing a single band from the moir\'e energy range and thus with two vacancies, only two moir\'e bands are left. In exploring the possible combinations of two vacancies, we find that this is a common pattern. However, we find that it is not quite universal. For example, the combination of two vacancies in the $DW$ region results in four bands in the moir\'e energy range, as shown in Fig.~\ref{fig:TPF-breaking}(b). The same result is also obtained when having two vacancies in the $AB$ region or one in each type of the $HC$/$LC$ site (not shown). Still, in all of these cases, the TPF is split. This non additivity of the effects of defects leads us to investigate the range of influence of each defect later in Sec.~\ref{sec:length-scales}. 

In order to further corroborate that the origin of TPFs in TBG is tied to a rank-$1$ perturbation, we next consider the case of a single but extended impurity. We model this by giving the impurity potential a Gaussian profile with a spread of $\sigma$ centered around the impurity site. We here set the strength of the Gaussian profile such that the perturbing potential is $1t$ in the central site. This can be regarded as a simple model for an impurity that affects multiple sites around its binding center, realistic for molecule adsorbates or for an adatom absorbed in the honeycomb lattice hollow site. For simplicity, we cut off the Gaussian at a distance of $2.5a$ away from the binding center, where $a$ is the graphene lattice constant, which we verify does not influence the results. In Fig.~\ref{fig:TPF-breaking}(c) we show the band structure resulting from such Gaussian impurity perturbation with the impurity center located in the $DW$ region, using $\sigma=0.8a$. We see that four bands remain mostly in the moir\'e energy range but notably the triple band crossing at $\mathbf{K}$ is clearly no longer present, meaning the TPF is split. We also verify that the splitting of the TPF holds for other impurity locations. In fact, when the extended impurity is centered in the $AA$ region we find that all degeneracies of the moir\'e bands at $\mathbf{K}$ are lifted, see Fig.~\ref{fig:TPF-breaking}(d). We also observe in this case a lifting of all four moir\'e bands, here illustrated for $\sigma=0.5a$. For a larger spread $\sigma=0.8a$ we find an even larger depletion as the moir\'e bands, which are lifted further in energy such that they leave behind only strongly dispersing states near $\mathbf{\Gamma}$ (not shown). This is a fascinating result showing how the entire moir\'e band structure can be destroyed by a single weak extended impurity. Taken together, our results in this subsection show that if the defects are not simple rank-$1$ perturbations, but instead, for example, multiple defects or extended impurities, then the TPF at $\mathbf{K}$ and $\mathbf{K'}$ is generally split. Based on these results, we do not expect TPF to be likely observed in TBG.

\subsection{Isolated defects and length scales}
\label{sec:length-scales}

In the previous subsections we explored the effects of a defect lattice with same periodicity as the pristine system, a model which can be implemented in practice by defect engineering. We now turn to the limit of an isolated defect. In the fully atomistic framework we use, this amounts to reducing the periodicity of the defect lattice, such that the distance between the different Bloch copies of the same defect become large enough as to not influence each other. We explore different supercell sizes and defect locations, with the constraint that all defects are within a single moir\'e unit cell inside the supercell, which we refer to as the defective moir\'e unit cell. In all defect configurations we explore, we find that an asymptotic behavior is reached in a specific spatial direction when the two defects are separated by three or more moir\'e unit cells along that particular direction. This means that for $3\times3$ or larger supercells single defects are effectively isolated from each other.

\begin{figure}[tb]
	\centering
	\includegraphics[width=\columnwidth]{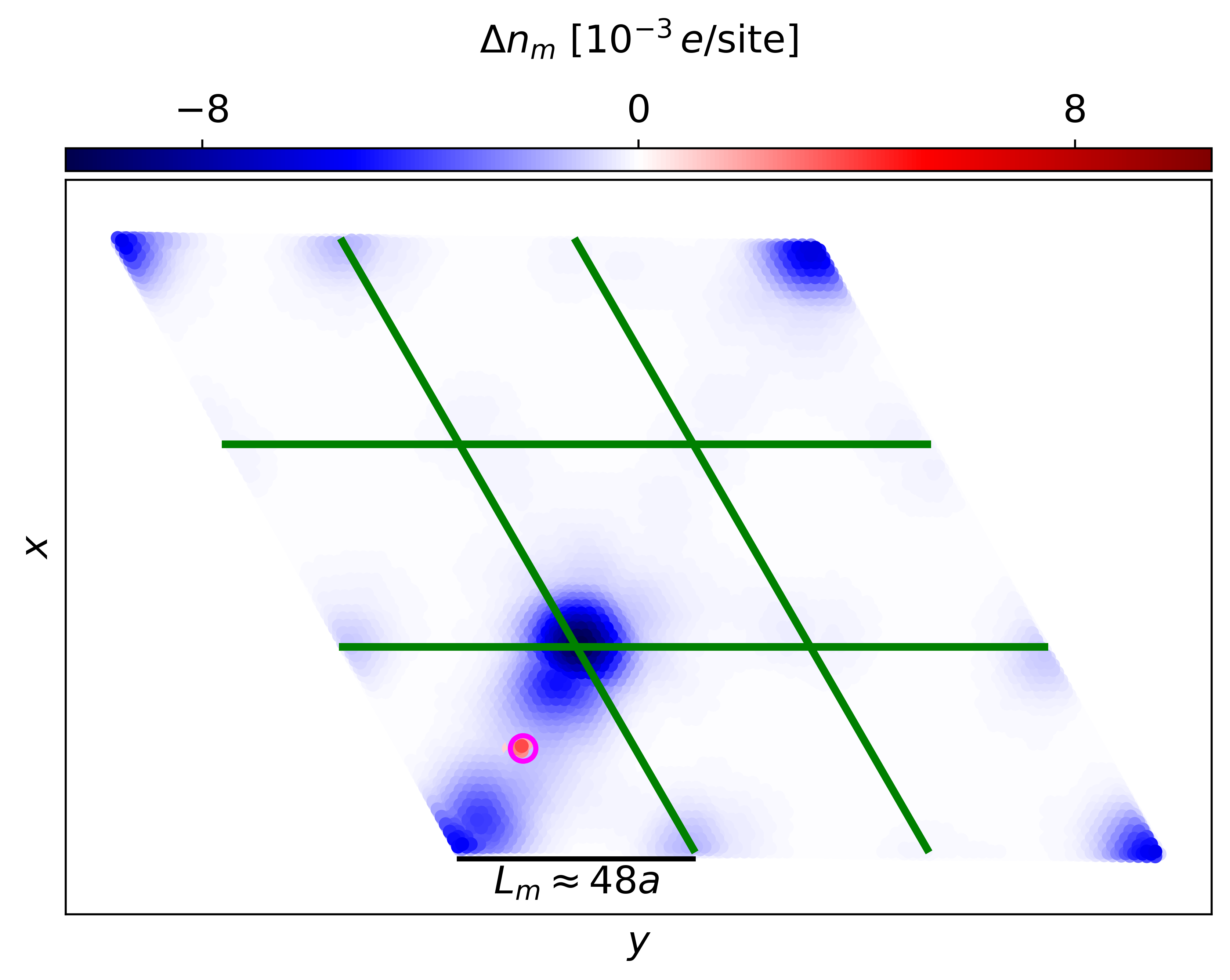}
	\caption{\label{fig:supercell} Change in moir\'e charge density $\Delta n_m$ for a single $DW$ vacancy in a $3\times3$ supercell. Depletion of the $AA$ regions neighboring the vacancy. Pink circle marks the vacancy location and green lines mark the individual moir\'e unit cells. The contributions from both layers are shown. Here $12\times12$ $k$-point sampling was used and $\theta \approx 1.2\degree$.}
\end{figure}

We illustrate the result for a single isolated vacancy in Fig.~\ref{fig:supercell}, where we plot the change in the moir\'e charge density $\Delta n_m$ for a $3\times 3$ supercell with a vacancy in a $DW$ region. As a guide to the eye we mark the boundaries between the individual moir\'e unit cells outlined by green lines. In the defective moir\'e unit cell (bottom left) we observe a similar charge density redistribution as in the $1\times 1$ supercell case discussed in Sec.~\ref{sec:defect-location}, with a clear depletion in the $AA$ regions closest to the vacancy. This can again be understood as one of the moir\'e bands associated with this unit cell leaving the moir\'e energy range. However, for a $3\times 3$ supercell, the band folding caused by the reduced periodicity results in $(3\times3)\times4=36$ bands in the moir\'e energy range in the pristine case. Thus, out of these $36$ bands, only one exits this energy range when the vacancy is introduced, causing a depletion of $1e$ distributed through the $AA$ region of the defective moir\'e unit cell. Away from the defective moir\'e unit cell we observe that the charge density quickly recovers to its pristine value, even in the $AA$ regions. We verify that a similar behavior is present also for larger supercells, which shows that we use a system size capable of modeling the asymptotic isolated defect limit.

\begin{figure}[tb]
	\centering
	\includegraphics[width=\columnwidth]{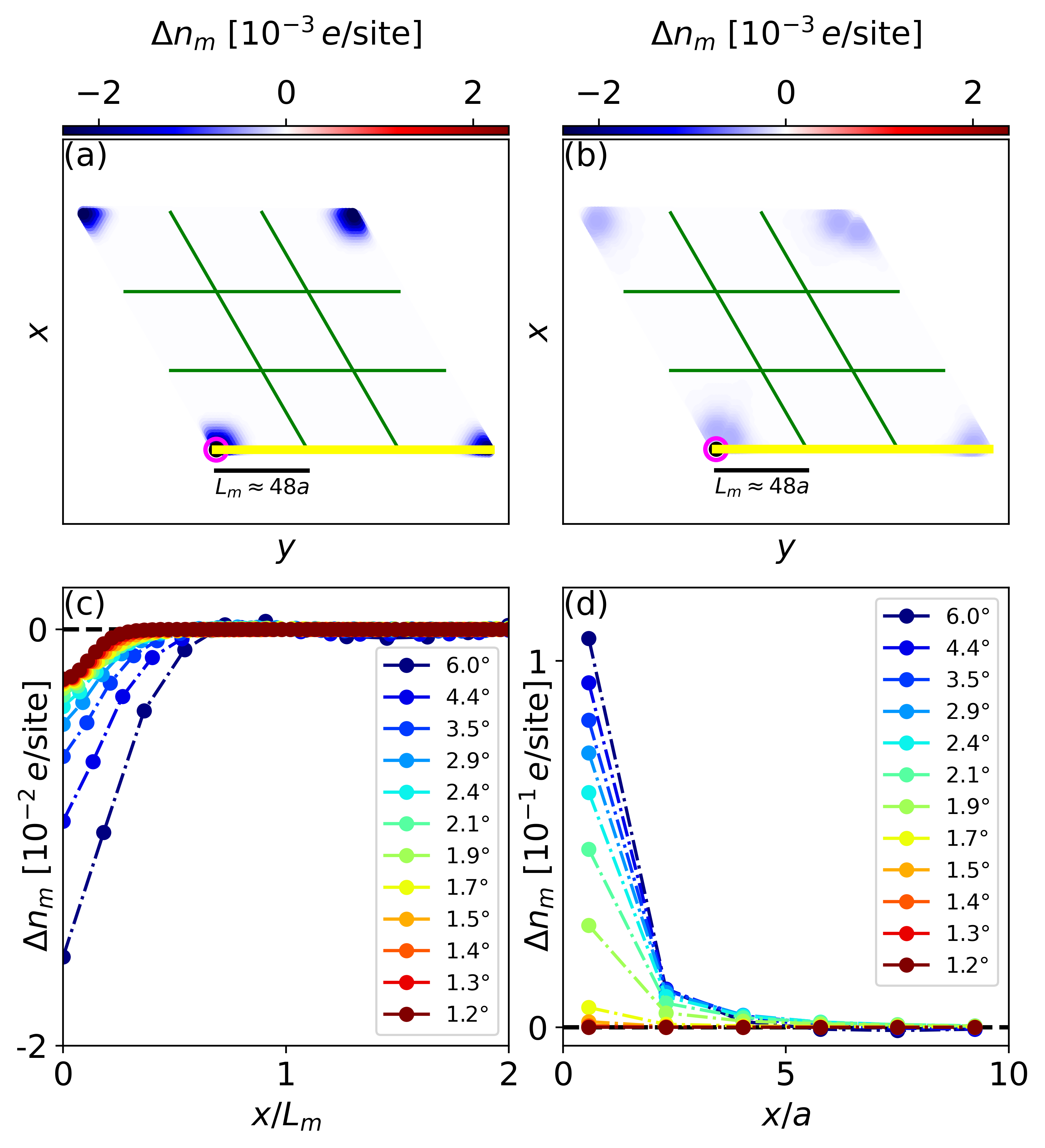}
	\caption{\label{fig:length-scales} Two length scales associated with defects in TBG. (a-b) Change in moir\'e charge density $\Delta n_m$ for a single $AA$ vacancy in the top layer in sublattice A at $\theta\approx1.2\degree$, i.e.~near the magic angle, for the $A$ (a) and $B$ (b) sublattice of the top layer. Green lines mark the individual moir\'e unit cells. (c-d) Change in moir\'e charge density $\Delta n_m$ along the line cut in (a,b) marked by the yellow line, for different twist angles, for the $A$ (c) and $B$ (d) sublattice of the top layer. The $x$-axis is scaled with respect to the moir\'e length $L_m$ (c) and atomic length $a$ (d), respectively. Here $6\times 6$ $k$-point sampling was used.}
\end{figure}

We find that the separation required between vacancies for reaching the isolated defect limit is reduced when the vacancy is in the $AA$ region. In this case, we find that even supercells as small as $2\times2$ give results converged to the isolated defect limit. We illustrate this in Fig.~\ref{fig:length-scales}(a,b), where we plot the contribution to $\Delta n_m$ on the sites of the top layer in sublattice $A$ and $B$, respectively, with the vacancy being on an $A$ sublattice site.  Note how only the $AA$ region with the vacancy has an altered charge density and thus the isolated defect limit is achieved already for $2\times2$ supercells. Moreover, it is clear that the charge density depletion is primarily in the same sublattice as the vacancy. For the vacancy-free layer we find a similar but smaller change in charge density in the $AA$ regions. Taken together, Figs.~\ref{fig:supercell} and \ref{fig:length-scales}(a,b) illustrate that a single vacancy generally influences the moir\'e pattern up to a distance of the moir\'e length $L_m$.

In order to further corroborate that the vacancies affect a region with radius of the order of $L_m$, we vary this length by changing the twist angle away from the magic angle, up to $\theta \approx 6 \degree$. We then study $\Delta n_m$ along a line cut passing through a vacancy using a $3\times3$ supercell, depicted by the yellow lines in Fig.~\ref{fig:length-scales}(a,b). This way, the vacancy concentration per unit area changes with twist angle, but the vacancy concentration per unit cell stays the same and can easily be compared. Also, the vacancies always stay isolated from each other and thus we stay within the isolated defect limit. We particularly choose a line cut direction that goes through a nearest neighbor site of the vacancy in the vacancy layer in order to also probe the graphene-like localized defect state that exists in this layer. The depletion of the $AA$ regions is however visible for any line cut direction. We further choose a vacancy location in the center-most site of the $AA$ region because this site is always present for all twist angles. This site is in sublattice $A$, while the layer index does not matter.

In Fig.~\ref{fig:length-scales}(c) we plot change in charge density for the $A$ sublattice sites along the line cut, where we know that graphene-like localized defect state does not contribute because it is primarily located in the opposite sublattice \cite{Pereira2006b, Ugeda2010}. Here we scale the $x$-axis with respect to the moir\'e length $L_m$, which is dependent on the varying twist angle. We observe a clear trend where the depletion in the $AA$ region recovers away from the vacancy with a length scale that approaches $L_m$, but is smaller especially for smaller angles approaching the magic angle. In contrast, in Fig.~\ref{fig:length-scales}(d) we show $\Delta n_m$ for the $B$ sublattice sites along the line cut. Here we let the $x$-axis be normalized with the atomic scale $a$, corresponding to the graphene lattice constant, as we find that the dominant contribution comes from the graphene-like localized defect state, giving a positive change in charge density on the atomic scale. This graphene-like localized defect state decays over a similar length scale, set by $a$, for all twist angles, further corroborating that its origin is due to graphene physics and not the moir\'e pattern. We thus conclude that isolated defects affect TBG on two different length scales. On the atomic length scale, $a$, it induces a localized defect state similar to that of in monolayer graphene. On the angle-dependent moir\'e length scale, $L_m$, it induces a strong charge depletion in the $AA$ regions, which near the magic angle can be understood from the removal of an entire moir\'e band from the low-energy region. We note that at larger angles, the moir\'e bands are not energetically separated from the conduction and valence bands, which means that the simple picture that one of the moir\'e bands leaves the moir\'e energy range breaks down. However, we verify the presence of a flat defect-induced band at the energy of the Dirac point in the vacancy limit, similar to the magic angle case, for all angles up to $6\degree$. The one exception we find to the latter behavior is for defects in $AB$-$LC$ sites, since in this case there is a moir\'e band replacement instead of a band removal. 

\section{Concluding remarks}
\label{sec:conclusions}

To summarize, in this work we show that the low-energy band structure of twisted bilayer graphene (TBG) is extremely sensitive to atomic size lattice defects even at very low concentrations. In particular, we show that a single weak non-magnetic impurity in each $AA$ region is able to cause a large depletion of charge in the low-energy regime and in the whole $AA$ region of the order of $1e$ per spin species due to the lifting of one of the low-energy moir\'e bands into the conduction bands. We investigate different defect locations and verify that this behavior is quite general and thus illustrates a direct way to manipulate the low-energy moir\'e band structure using impurities. The only notable exception we find is for a special set of defect sites in the $AB$ region, where a band replacement process happens instead, where one moir\'e band is lifted to the conduction band, while another joins from the valence band, resulting in a reconstruction of the original low-energy band structure in the vacancy limit. We strongly suspect that this band replacement directly influences the topology of the moir\'e band structure although that remains to be verified. 

We further find that the previously reported defect-induced TPFs in TBG \cite{Ramires2019b}, which represent a triple degeneracy at the Dirac point, rely on the rank-$1$ perturbation characteristic of single-site defects, and is thus not generally present. In fact, we show that the introduction of multiple defects or more realistic extended impurities easily split the TPF by lifting the degeneracy at the Dirac point, in some cases even completely removing the Dirac point. Finally, we use supercells to reach the isolated defect limit, where we find the previous results to hold locally in the moir\'e unit cells surrounding the defect. By varying the twist angle we are further able to identify two length scales, with the atomic scale displaying a graphene-like localized defect state, and the twist angle-dependent moir\'e length controlling the charge depletion of the $AA$ region and accompanied moir\'e band restructuring.

Our results establish how non-magnetic impurities and vacancies drastically change the band structure and charge density of TBG at and near the magic angle, which can be experimentally verified with ARPES, STM, or transport measurements. Incorporating these profound changes of the moir\'e bands will further be important for analyses of quasiparticle interference patterns. These measurements should be performed at low enough temperatures for good energy resolution (we estimate around $\sim10$K), but above the critical temperatures of any emerging electronic orders of TBG. We expect that the changes induced by defects will also have profound impact on these electronic orders, including the superconducting and correlated insulator orders, since these orders depend not only on the interactions, but also heavily on the underlying normal-state band structure. 

This impact will be particularly large on any mechanism relying on the symmetries of the system or number of moir\'e bands, as defects strongly modify the low-energy moir\'e band structure through band lifting and band replacement processes. In fact, we show that the number of moir\'e bands can easily change from the pristine case of four to three or two, or even be completely annihilated, with only strongly dispersive bands left in the low-energy region. This sensitivity of TBG with respect to impurities and vacancies demonstrate the need to understand the disorder level before further analyzing any electronic ordered state. It also opens up the possibility of using defects to engineer the low-energy electronic structure of TBG in order to produce a desired number of flat bands and thereby possibly other electronic orders.

\section*{Acknowledgments}
\label{sec:acknowledgments}
We acknowledge financial support from the Swedish Research Council (Vetenskapsr\aa det) grant no.~2018-03488 and the Knut and Alice Wallenberg Foundation through the Wallenberg Academy Fellows program. Computations were enabled by resources provided by the National Academic Infrastructure for Supercomputing in Sweden (NAISS) and the Swedish National Infrastructure for Computing (SNIC) at the computing center UPPMAX, partially funded by the Swedish Research Council through grant agreements no.~2022-06725 and no.~2018-05973.

\appendix
\section{Defects in $DW$ and $AB$-$HC$ sites}
\label{app:defect-location}

\begin{figure}[htb]
	\centering
	\includegraphics[width=\columnwidth]{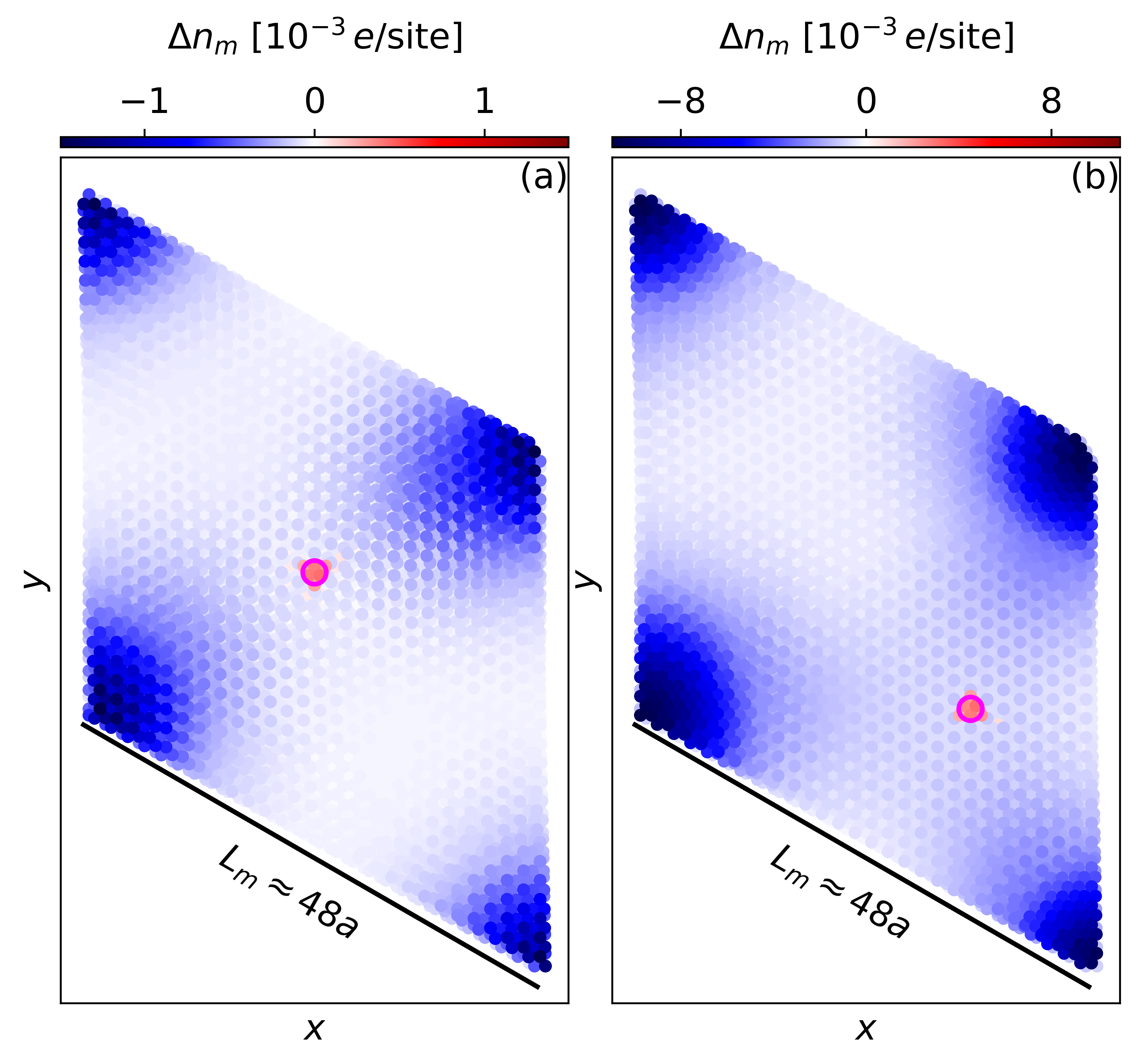}
	\caption{\label{fig:charge-density-extra} Change in moir\'e charge density $\Delta n_m$ from pristine magic-angle TBG due to a vacancy in the $DW$ region (a) and in an $AB$-$HC$ site (defined in main text) (b), highlighting the strong charge depletion of the $AA$ region due to the removal of one of the moir\'e bands, and also a smaller depletion in the larger vicinity of the defect. Pink circles mark the vacancy site and $k$-points were sampled in a $12\times12$ grid. In each panel the contributions from both layers are shown. Here $\theta \approx 1.2\degree$.}
\end{figure}

In the main text we discuss the effects of defects in $AA$ and $AB$-$LC$ sites extensively, while focusing less on defects in the $DW$ and $AB$-$HC$. The reason for this is that these latter sites show a behavior quite similar to that of $AA$ sites. Still, for completeness, we here provide relevant data for defects in the $DW$ and $AB$-$HC$ sites, supporting the conclusions in the main text. In Fig.~\ref{fig:charge-density-extra}(a,b) we show the change in moir\'e charge density $\Delta n_m$ induced by a vacancy in $DW$ and $AB$-$HC$ sites, respectively, using a $1\times 1$ supercell at $\theta \approx 1.2\degree$ and with the vacancy highlighted by a pink circle. Most importantly, and as mentioned in the main text, we find that the introduction of a defect induces a strong depletion of the $AA$ region even in these cases where the defect is located far away from it. We additionally find a slight depletion of states in an extended region between the defect and the nearest $AA$ regions. This provides further evidence of the effect that atomic size defects have on the moir\'e scale.

\begin{figure}[h!]
	\centering
	\includegraphics[width=\columnwidth]{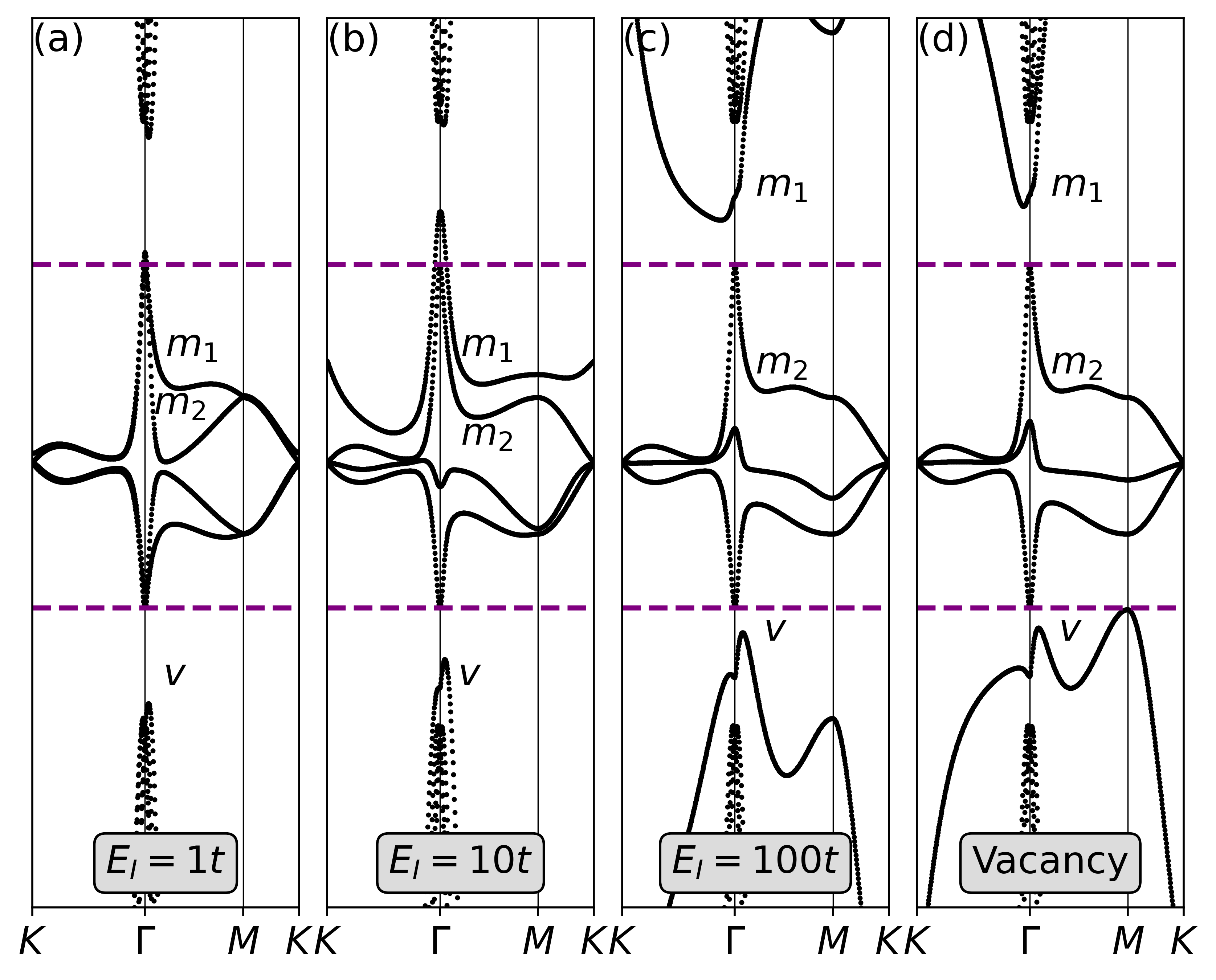}
	\caption{\label{fig:band-evolution-DW} Band removal process for a defect in the $DW$ region. (a-d) Band structure for potential impurities with strengths of $1t$, $10t$, $100t$, and a vacancy, respectively. Here $\theta \approx 1.2\degree$.}
\end{figure}

\begin{figure}[h!]
	\centering
	\includegraphics[width=\columnwidth]{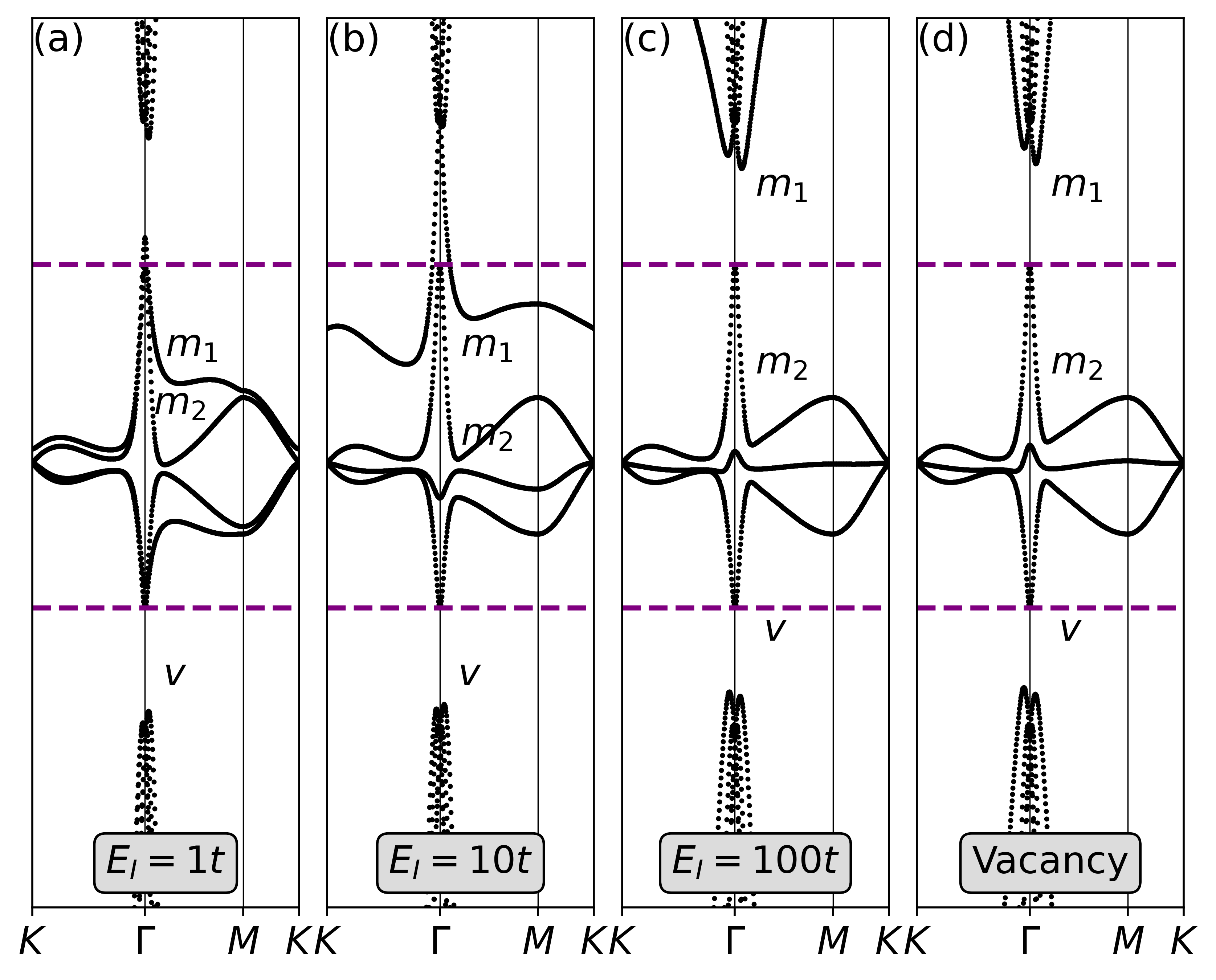}
	\caption{\label{fig:band-evolution-ABHC} Band removal process for a defect in an $AB$-$HC$ site. (a-d) Band structure for potential impurities with strengths of $1t$, $10t$, $100t$, and a vacancy, respectively. Here $\theta \approx 1.2\degree$.}
\end{figure}

Once again, the depletion of the $AA$ regions can be understood from the evolution of the band structure as we interpolate between the pristine and vacancy limits with an impurity of finite strength, as we illustrate in Figs.~\ref{fig:band-evolution-DW} and \ref{fig:band-evolution-ABHC} for $DW$ and $AB$-$HC$ impurities, respectively. In both cases we observe a band removal process akin to the the one discussed in the main text, with the $m_1$ band leaving the moir\'e energy range and joining the conduction bands. The impurity energies required to trigger the removal process is higher than for an $AA$ impurity, with over $10t$ being required to remove the $m_1$ band from the moir\'e energy range. Just as in the $AA$ defect case, this explains the charge depletion of the $AA$ region, since it is there the moir\'e bands are located. We also note that the degeneracy of the moir\'e bands at the $\Gamma$-point, discussed in Sec.~\ref{sec:defect-location}, is lifted, since the defect site in these cases is not on the rotation axis of a $C3$ symmetry.

\bibliography{bibliography}

\end{document}